\documentclass[preprint,prb,showpacs,preprintnumbers,amsmath,amssymb]{revtex4}

\usepackage{graphicx}
\usepackage{dcolumn}
\usepackage{bm}

\begin{document}
\draft
\title
{
Negative Hall coefficients of heavily overdoped La${}_{2-x}$Sr${}_x$CuO${}_4$
}

\author{I. Tsukada}
\email{ichiro@criepi.denken.or.jp}
\author{S. Ono}
\affiliation{Central Research Institute of Electric Power Industry, 
2-11-1 Iwadokita, Komae-shi, Tokyo 201-8511, Japan}

\date{\today}

\begin{abstract}
The Hall coefficient ($R_{\rm H}$) is investigated along the in-plane 
direction for carefully prepared La${}_{2-x}$Sr${}_x$CuO${}_4$ thin films 
from $x$=0.08 to $x$=0.40. 
It is found that $R_{\rm H}$ becomes almost temperature independent 
around room temperature for $x \geq$ 0.24, 
and its sign smoothly changes from positive to negative between $x$ = 0.28 
and 0.32, showing contrasting behavior to previously reported results. 
This result indicates that the carrier can be doped smoothly 
across the hole- to electron-dominant regions. 
According to the sign change of $R_{\rm H}$, superconductivity disappears 
for $x \geq$ 0.32 films, which suggests that a positive $R_{\rm H}$ 
is one of the necessary conditions for superconductivity 
in $p$-type cuprates. 
\end{abstract}

\maketitle


Evolution of superconductivity according to carrier doping is an essential 
feature of high-temperature (high-$T_{\rm c}$) cuprate superconductors. 
All high-$T_c$ cuprates have a two-dimensional square-lattice 
CuO${}_2$ plane in common, 
on which an electronic ground state drastically changes from insulator 
to doped semiconductor, superconductor, and normal metal.
\cite{Takagi1} 
One of the key concepts needed to reveal the mechanism 
of high-$T_{\rm c}$ superconductivity is understanding 
how superconductivity occurs by carrier doping. 
For that purpose, people have started extensive studies on the carrier 
doping effect in the underdoped region soon 
after the discovery of high-$T_c$ cuprates. 
Due to an incredible amount of studies, we have now come to believe 
that a strong coupling among charge, spin, and lattice plays 
a significant role in the physical properties of underdoped cuprates, 
even though its direct relation to the occurrence of superconductivity 
is still veiled. 
On the other hand, physical properties of the overdoped metallic region 
have not been fully invetigated mainly due to difficulty 
in preparing high-quality samples. 
Even for well-studied La${}_{2-x}$Sr${}_x$CuO${}_4$ (LSCO), for example, 
it has been known that the oxygen deficiency becomes significant 
in the overdoped side, typically when $x$ exceeds 0.28.
\cite{Torrance1}
Thus, the experimental approach to superconductivity from the overdoped side 
using high-quality single crystals or thin films is limited only 
to two systems: LSCO,
\cite{Takagi1,Takagi2,Nakamae1} 
and Tl${}_2$Ba${}_2$CuO${}_{6+\delta}$ (Tl2201),
\cite{Manako1,Mackenzie1} 
and, because of the toxicity of Tl, LSCO has been and is still 
the most important compound for the study 
from the {\it nonsuperconducting} overdoped side.

It is still an open question whether or not heavily overdoped 
LSCO is really a normal Fermi liquid. 
Takagi {\it et al.}
\cite{Takagi2} 
reported for $x$ = 0.34 single-crystal thin films that 
in-plane resistivity ($\rho$) does not follow simple $T^2$ behavior 
but is well fitted by $T^{1.5}$.
Nakamae {\it et al.} 
\cite{Nakamae1} 
has recently found ideal $T^2$ resistivity behavior below $\sim$ 55~K 
for $x$ = 0.30 in a single crystal, but the resistivity 
at higher temperature region follows $T^{1.6}$ behavior. 
Similar temperature dependence has been reported for overdoped 
Tl2201 single crystals.
\cite{Manako1,Mackenzie1} 
The situation mentioned above means that there is a difficulty in judging 
whether or not heavily overdoped high-$T_c$ cuprates merge 
into a normal Fermi liquid only from a resistivity measurement.

The Hall coefficient ($R_{\rm H}$) is a more robust parameter 
than resistivity, and is used to obtain direct information of carriers. 
However, the $R{}_{\rm H}$ data for heavily overdoped LSCO has still been 
controversial. 
The first systematic study on the Hall effect of LSCO was done 
by Takagi {\it et al.}
\cite{Takagi1} 
immediately after the discovery of high-$T_{\rm c}$ cuprates.
They reported for polycrystalline LSCO that $R_{\rm H}$ possibly becomes 
negative at $x$ = 0.32 and 0.34. 
However, they also described in the same paper that 
an $x$ = 0.34 single crystal shows a positive $R_{\rm H}$. 
Such an unclear situation was reported also by Hwang {\it et al.}
\cite{Hwang1} 
for polycrystalline sintered samples and single-crystal thin-film samples. 
They reported that the room-temperature $R_{\rm H}$ becomes almost zero 
both for $x$ = 0.30, and 0.34; 
$R_{\rm H}$ does not obviously go to a negative side, 
but seems to stay around zero. 
Suzuki
\cite{Suzuki1} 
has reported the $R_{\rm H}$ for LSCO thin films prepared by sputtering; 
he found that the $x$ = 0.36 sample shows a positive $R_{\rm H}$, 
which is clearly inconsistent with the above two results.
\cite{Takagi1,Hwang1} 
These three results suggest the presence of two hurdles 
when we study the Hall effect of heavily overdoped LSCO. 
The first hurdle is that we must be very careful about the sample 
misalignment/orientation. 
In particular, we need to align the current direction perfectly parallel to the 
CuO${}_2$ plane, because even a slight misalignment may seriously influence 
the sign of Hall coefficients. 
As was demonstrated by Tamasaku {\it et al.},
\cite{Tamasaku1} 
for the $x$ = 0.30 crystal, 
$R_{\rm H}$ is positive for $i \parallel$ CuO${}_2$ plane 
while it is  negative for $i \perp$ CuO${}_2$ plane.
Thus it becomes very important to minimize the misalignment 
between the CuO${}_2$ plane direction and the current path direction 
as much as possible, 
especially when the magnitude of $R_{\rm H}$ is very small as 
in the overdoped samples. 
The other hurdle is the oxygen deficiency as was mentioned before.
\cite{Torrance1,Kanai1} 
It is partially helpful to anneal the samples for a very long time 
under high-oxygen pressure.
\cite{Torrance1,Takagi2,Nakamae1} 
However, high-pressure annealing is not possible everywhere, 
and is hardly applicable to thin-film samples.

In this paper, we propose to solve these two problems by using 
epitaxial thin films combined with a strong oxidation technique, 
and carry out Hall-effect measurements to demonstrate that a sufficient 
amount of holes can be doped in LSCO at least up to $x$=0.40. 
We pay attention to the following two points in particular; 
a) whether $R_{\rm H}$ indeed becomes negative above a certain $x$, 
or just stays at zero, and 
b) how the temperature dependence of $R_{\rm H}$ goes with increasing $x$, 
or in other words, 
whether an ordinary Fermi liquid ($T$-independent $R_{\rm H}$) indeed 
can be observed in the heavily overdoped LSCO.

\begin{figure}
\begin{center}
\includegraphics*[width=100mm]{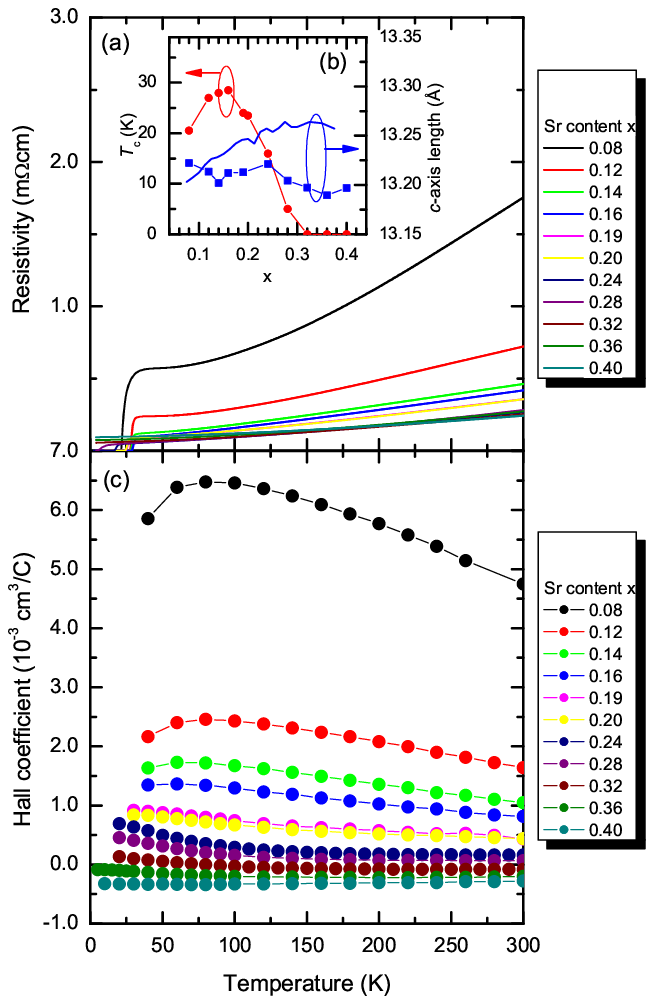}
\end{center}
\caption{
(color online). 
(a) Temperature dependence of {\it dc} resistivity for 
0.08 $\leq  x \leq$ 0.40. 
(b) The line with red filled circles shows the $x$ dependence 
of $T_c$ (left axis). It shows a smooth change with $x$. 
The line with blue filled squares shows the $x$ dependence of the $c$-axis length 
which shows shorter values than those of bulk crystals (blue solid line)
\protect\cite{Radaelli1} for $x \geq$ 0.12. 
(c) Temperature dependence of $R_{\rm H}$. 
The data were taken by sweeping the magnetic field between $\pm$1~T 
for 0.08 $\leq x \leq$ 0.22, and $\pm$6~T for 0.24 $\leq x \leq$ 0.40.
}
\label{Fig1}
\end{figure}

All films were grown by a pulsed laser deposition (PLD) technique 
with low-pressure pure ozone as an oxidant.
\cite{Tsukada1} 
The films were all highly $c$-axis oriented as confirmed 
by x-ray diffraction. 
We used SrTiO${}_3$ (100) as a substrate in order to not only compare 
our data with several published data but also decrease the influence 
of a strong epitaxial-strain effect. 
\cite{Tsukada2}
It has been known that LSCO is highly susceptible to the lattice mismatch, 
and it is almost impossible to perfectly reproduce the transport properties 
of bulk crystals in thin-film samples.
\cite{Sato1,Locquet1,Bozovic1} 
However, it has been experimentally shown
\cite{Tsukada2} 
that the epitaxial strain to LSCO is not as large on SrTiO${}_3$ (100) 
as on LaSrAlO${}_4$ (001) and LSAT (100). 
We actually confirm that the effect of epitaxial strain is not 
killed completely even using SrTiO${}_3$ (100), 
and the present films indeed have shorter $c$ axes 
than the bulk crystals have
\cite{Radaelli1} 
for almost the entire $x$ range (Fig.~\ref{Fig1}(b)), 
which indicates that finite tensile strain survives in our films. 
Substrate temperature during the deposition was set around 
820 - 840${}^{\circ}$C, 
and the film thickness was kept around 1500 - 2500~{\AA}. 
Samples are separated into two groups; 
the films for $x$ = 0.08 - 0.22 were processed into a small 
six-terminal shape using photolithography and wet etching,
\cite{Balakirev1} 
while those for $x$ = 0.24 - 0.40 were prepared using a metal mask 
(six terminal shape) in order to omit any post heat treatments. 
Ozone is indispensable for two reasons; 
first, we can keep the chemical composition of the films 
identical to the targets. 
It is known that in a usual PLD or sputtering technique that the excess Cu 
is necessary to be added in the target in order to compensate 
the Cu deficiency in the films.
\cite{Suzuki1} 
However, this procedure makes the relation between Sr concentration and real 
hole density unclear and uncontrollable. 
Because the reduction of gas pressure can make the mean-free path 
of each ablated species longer than the target-substrate distance, 
the deviation of chemical composition is automatically suppressed as a result.
\cite{Tsukada3} 
We set the ozone pressure to 10~mPa during the film deposition, 
and increase it to 15~mPa during the cooling down procedure. 
Second, we can utilize a stronger oxidation ability of ozone 
than high-pressure oxygen. 
It is known that the activity of ozone at $T$ = 300~K reaches 
10${}^{19}$ times larger than that of oxygen,
\cite{Suzuki2} 
implying an efficient compensation of oxygen deficiency. 
In the present case, we suppress the oxygen deficiency by supplying ozone 
until the film temperature decreases by 60 $\sim$ 65${}^{\circ}$C. 
This procedure is applied for the sample with 0.20 $\leq x \leq$ 0.40. 
The other films with lower $x$ are annealed after the growth 
under the same conditions applied to bulk single crystals.
\cite{Komiya1}


The temperature dependence of the in-plane resistivity of all the films 
is summarized in Fig.~\ref{Fig1}(a). 
The figure shows a systematic change in magnitude and $T_c$. 
The magnitude of the resistivity in moderately doped films is slightly higher 
than that of bulk single crystals,
\cite{Ando1} 
which is frequently observed when films are grown on SrTiO${}_3$,
\cite{Suzuki1,Sato1,Locquet2} 
and considered to be one of the effects of the remanent epitaxial strain. 
The zero resistivity temperature ($T_{\rm c0}$) is plotted as functions 
of $x$ in Fig.\ref{Fig1}(b), 
in which a conventional dome-shaped curvature is properly reproduced. 
The temperature dependence of Hall coefficients for these films are 
shown in Fig.~\ref{Fig1}(c). 
In contrast to the resistivity data, $R_{\rm H}$ shows better coincidence 
for $x \leq$ 0.2 with those of bulk single crystals;
\cite{Ando1} 
$R_{\rm H}$ moderately increases towards low temperatures, 
shows a peak between 50 - 100~K, and rapidly decreases towards 
superconducting transition. 
The room temperature values are plotted in the inset of Fig.~\ref{Fig3}, 
which shows a smooth evolution of low-field $R_{\rm H}$ by Sr doping.

\begin{figure}
\begin{center}
\includegraphics*[width=150mm]{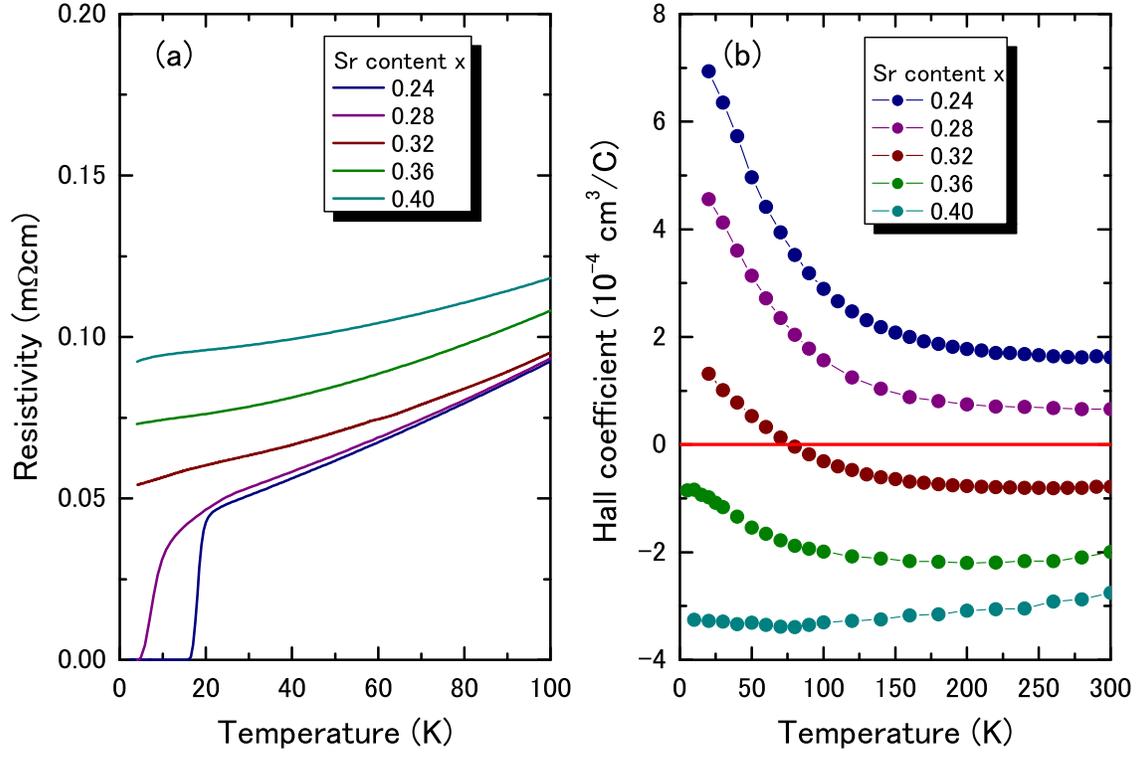}
\end{center}
\caption{
(color online). 
Close-up data of resistivity (a) and $R_{\rm H}$ (b) 
for $x \geq$ 0.24. 
}
\label{Fig2}
\end{figure}
\begin{figure}
\begin{center}
\includegraphics*[width=150mm]{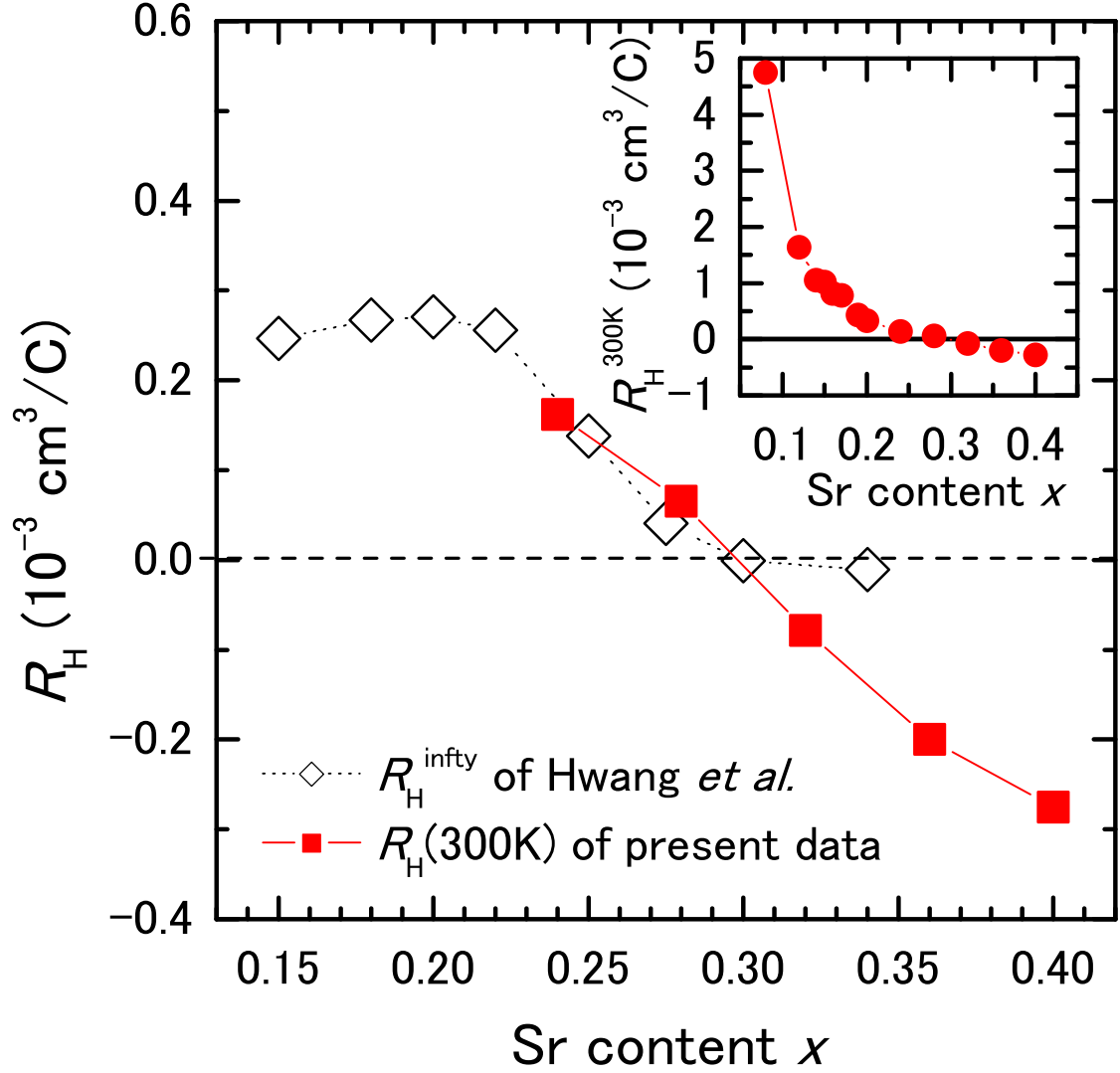}
\end{center}
\caption{
(color online). 
Plot of $R_{\rm H}$ at 300K of the present work for $x \geq$ 0.24 
with $R^{\infty}_{\rm H}$ from Ref. 7. 
Inset shows the $x$ dependence of $R_{\rm H}$ at 300K 
for the entire doping range. 
}
\label{Fig3}
\end{figure}

We focus on the heavily overdoped region in more detail. 
We again plot the resistivity and Hall data for $x \geq$ 0.28 
in Figs.~\ref{Fig2}(a) and (b). 
As is easily seen, superconductivity is clearly observed for $x$ = 0.24 and 0.28, 
while zero resistivity is no longer observed for $x \geq$ 0.32, 
which is consistent with Torrance {\it et al.}
\cite{Torrance1} 
The transition width of $x$ = 0.28 is far wider than 
that of $x$ = 0.24, which indicates that the oxygen distribution becomes 
inhomogeneous with increasing $x$. 
The onset of superconductivity observed in the $x$ = 0.32 film 
may be caused also by the oxy1gen inhomogeneity. 
It is intriguing that the room-temperature $R_{\rm H}$ changes its sign 
in accordance with the disappearance of superconductivity. 
For $x$ = 0.28, $R_{\rm H}$ keeps a very small but positive value 
of the order of 10${}^{-5}$~cm${}^3$/C from 300~K to 200~K, 
and begins increasing towards the lower temperatures. 
The boundary between positive and negative $R_{\rm H}$ is found 
between $x$ = 0.28 and 0.32. 
For $x$ = 0.32, $R_{\rm H}$(300K) is no longer positive but its temperature 
dependence looks similar to that of the $x$ = 0.28 film, 
and also to that reported for $x$ = 0.30 bulk single crystals.
\cite{Tamasaku1} 
What is newly discovered is the $R_{\rm H}$ behavior of 
$x$ = 0.36 and 0.40 films. 
For $x$ = 0.36, $R_{\rm H}$(300K) increases its magnitude 
to the negative side, is no longer positive down to $T$ = 0~K, 
and $dR_{\rm H} / dT$ turns positive at room temperature 
suggesting that the high-temperature limit of $R_{\rm H}$ is 
much closer to zero. 
To our knowledge, these features have never been reported for LSCO. 
For $x$ = 0.40, the low-temperature upturn behavior is no longer observed 
and $R_{\rm H}$ behaves as a typical metal. 
The doping dependence of $R_{\rm H}(T)$ behavior reminds us 
of the doping dependence of $n$-type cuprates 
(Nd${}_{2-x}$Ce${}_x$)CuO${}_4$ (NCCO) and 
(Pr${}_{2-x}$Ce${}_x$)CuO${}_4$ (PCCO)
\cite{Wang1,Jiang1,Fournier1} 
in which $R_{\rm H}$ changes its sign from negative to positive 
with increasing Ce${}^{4+}$ substitutions for Nd${}^{3+}$ or Pr${}^{3+}$. 
Therefore, the present data strongly indicates the similarity of 
electronic states between LSCO and NCCO (PCCO) at heavily overdoped regions.

It should be also noted that the occurrence of superconductivity 
and the sign change of $R_{\rm H}$ is not always correlated. 
We should remember that overdoped nonsuperconducting Tl2201 has a positive 
$R_{\rm H}$,
\cite{Manako1} 
which is one of the counterevidences. 
Moreover, we have recently revealed that strong epitaxial strain can kill 
superconductivity completely at any doping region,
\cite{Hanawa1} 
which also indicates that superconductivity can disappear 
even when $R_{\rm H} >$ 0. 
To reveal how the epitaxial strain influences the sign of $R_{\rm H}$ 
is our next subject.


Let us compare the present negative Hall coefficients with 
previously published data. 
Hwang {\it et al.}
\cite{Hwang1} 
analyzed the temperature dependence of the Hall coefficient 
of LSCO with 0.15 $\leq x \leq$ 0.34 using a scaling function expressed as 
$R_{\rm H}(T) = R^{\infty}_{\rm H} + R^{\ast}_{\rm H} f(T/T^{\ast})$. 
In this manner, $R^{\infty}_{\rm H}$ gives the value of the high-temperature 
$T$-independent part of $R{}_{\rm H}$, and in our case 
$R_{\rm H}$(300~K) for $x \geq$ 0.24 is a good approximation 
of $R^{\infty}_{\rm H}$. 
Figure~\ref{Fig3} shows the $x$ dependence of $R_{\rm H}$(300~K) of our films 
with $R^{\infty}_{\rm H}$ shown in the inset of Fig.~2 of Ref.~7. 
In contrast to Hwang's data, our $R_{\rm H}$(300~K) data shows no saturation 
even above $x$ = 0.30; 
it keeps decreasing to a negative side while keeping 
the gradient, $dR_{\rm H} / dx$, unchanged. 
This result strongly indicates that the carrier doping is doned 
smoothly across the hole- to electron-dominant regions. 
It should be noted that this smooth sign change is consistent 
with the theoretical prediction by Shastry {\it et al.},
\cite{Shastry1} 
in which the $ac$ Hall coefficients were proposed to be a better 
measure of carrier concentration. 
If our $R_{\rm H}$ around 300K can be regarded as an approximation 
of $ac$ Hall coefficients at a sufficiently high-frequency limit, 
our data looks consistent with these predictions in the following two points; 
1) the sign change occurs at $x$ = 0.30 $\sim$ 0.35, and 
2) the absence of saturating behavior at the sign-change concentration. 
The smooth sign change at similar Sr content was also predicted 
in a different theory constructed by Stanescu and Phillips,
\cite{Stanescu1} 
which again supports that the smooth sign change is a natural consequence 
of the $p$-type doping in LSCO.

\begin{figure}
\begin{center}
\includegraphics*[width=150mm]{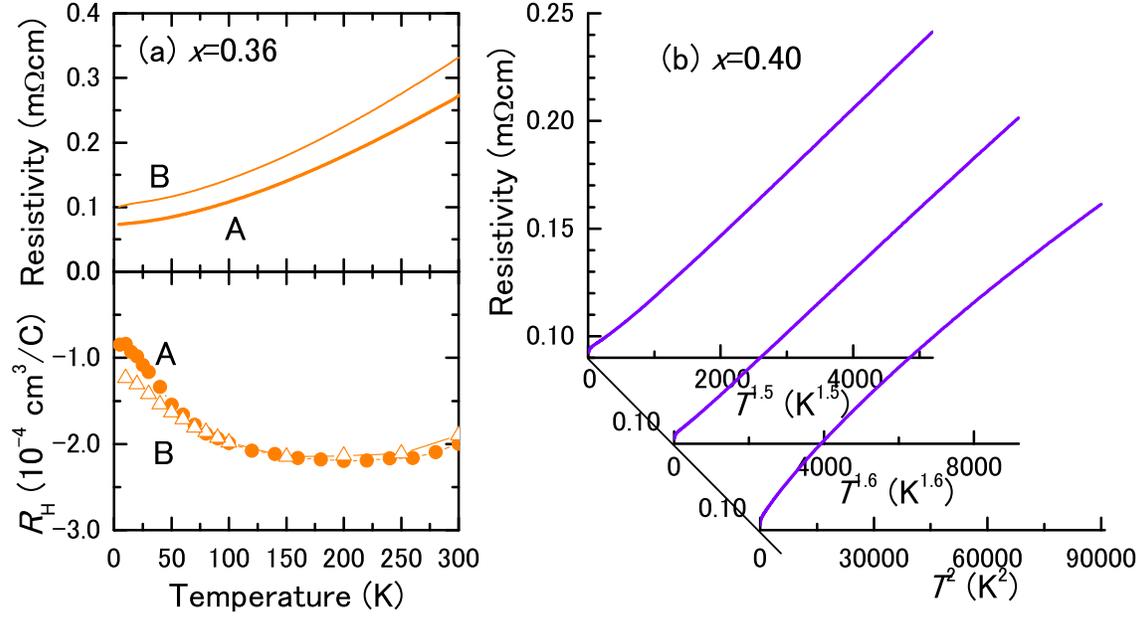}
\end{center}
\caption{
(color online). 
(a) Temperature dependence of resistivity and Hall coefficients 
for the two $x$ = 0.36 films. 
(b) $T^{1.5}$, $T^{1.6}$, and $T^2$ plots of the resistivity data for $x$ = 0.40. 
}
\label{Fig4}
\end{figure}

Comparison with the results of ARPES measurements is also intriguing. 
Doping dependence of the Fermi surface in LSCO has been systematically 
investigated by Ino {\it et al.}
\cite{Ino1} 
According to them, the Fermi surface turned to an electron-like shape 
at $x$ = 0.30, 
{\it i.e.,} closed the Fermi surface around the $\Gamma$ point. 
The valence band clearly crosses the Fermi energy between the (0,0) and 
($\pi$, 0) points (antinodal direction), 
and the Fermi surface approaches that of conventional metal. 
This is roughly consistent with the negative Hall coefficients in our films. 
One may wonder if the consistency between $R_{\rm H}$ and ARPES data is 
invalid because the latest ARPES measurements by Yoshida {\it et al.}
\cite{Yoshida1} 
reveal that even at $x$ = 0.22 the Fermi surface has a closed shape. 
However, we should remember that the sign of $R_{\rm H}$ is not determined 
simply by a topology of the Fermi surface. 
It is widely known based on Ong's argument
\cite{Ong1} 
that $R_{\rm H}$ is mainly determined by the portion of the convex and 
concave parts of the Fermi surface, 
and the Fermi surface of the $x$ = 0.22 crystal
\cite{Yoshida1} 
mostly consists of the convex part, which is consistent with positive $R_{\rm H}$ 
at $x$ = 0.22. 
Kontani {\it et al.}
\cite{Kontani1} 
calculated $R_{\rm H}$ more rigorously. 
According to them, the vertex correction (back-flow process) should be 
taken into account when calculating $R_{\rm H}$, 
and the portion of the Fermi surface inside and ouside 
the magnetic Brillouin zone is important. 
Even according to their conclusion, ARPES data of the $x$ = 0.22 
single crystal looks consistent with positive $R_{\rm H}$. 
One may suspect that the ARPES data was taken at $T$ = 20~K, and hence 
it should be compared with the $R_{\rm H}$ also at $T$ = 20~K. 
However, we think that the low-temperature upturn of $R_{\rm H}$ is not 
a good measure of hole concentration, because it is strongly influenced 
by impurity scattering. 
To give an example, we show resistivity and Hall data for two different 
$x$ = 0.36 films in Fig.~\ref{Fig4}(a). 
One can see that a lower resistivity sample (A) shows a stronger upturn 
in $R_{\rm H}$ at low temperatures, 
while the magnitude of $R_{\rm H}$ at room temperatures 
are insensitive to a difference of resistivity. 
Thus, we may regard the room-temperature $R_{\rm H}$ as a more direct 
measure of carrier concentration.

After seeing the increase of $|R{}_{\rm H}|$ to the negative side, 
we may reasonably understand the increase of resistivity 
when $x$ exceeds 0.32 to be intrinsic. 
In previous studies, the oxygen deficiency that becomes remarkable 
in heavily overdoped region is believed to be the main reason 
of this resistivity increase.
\cite{Takagi1,Torrance1,Ando1}
The present result indicates that the Fermi surface starts 
to shrink once the doping exceeds $x$ $\approx$ 0.3, 
which leads to the decrease of in electron concentration. 
Thus, the resistivity may increase if the effective mass of electron 
is kept unchanged. 
It is interesting to perform an optical reflectivity measurement, 
with which we can estimate the density of the free carriers 
summed up to a finite frequency.


We briefly mention the temperature dependence of resistivity 
of the $x$ = 0.40 film. 
Figure~\ref{Fig2}(b) suggests an absence of special Sr concentration 
at which $R_{\rm H}$ strictly becomes temperature independent. 
However, the $R_{\rm H}$ for $x$ = 0.40 shows weaker temperature dependence 
than the others. 
Thus, it is meaningful to analyze the temperature dependence 
of the resistivity of this film. 
The result is shown in Fig.~\ref{Fig4}(b). 
We do not observe perfect $T^2$ behavior as similar to previous results.
\cite{Takagi2} 
The temperature dependence is better expressed as $T^{1.5} \sim T^{1.6}$, 
which indicates the mixture of $T$ linear behavior with the $T^2$ one. 
We don't know whether the further doping realizes pure $T^2$ behavior or not. 
However, the presence of $T$-linear term in $x$ = 0.40, where $R_{\rm H}$ 
becomes almost temperature independent, indicates that the electron-electron 
scattering is not the only scattering process even in heavily overdoped LSCO.


In summary, we measured the Hall coefficients of LSCO thin films grown under 
ozone atmosphere, which effectively suppresses the oxygen deficiency. 
These films demonstrate a smooth change of the Hall coefficients 
from positive to negative at $T$ = 300K, which shows a clear 
contrast to the previously reported results. 
The temperature dependence of Hall coefficients almost diminishes 
at $x$ = 0.40 while the temperature dependence of resistivity still 
follows $T^{1.5} \sim T^{1.6}$. 
The sign change of $R{}_{\rm H}$ is consistent with the shape of 
the Fermi surface determined by an ARPES measurement.

We appreciate Yoichi Ando, Seiki Komiya, F. F. Balakirev, G. S. Boebinger, 
H. D. Drew, H. Sato, H. Kontani, A. Ino, P. Phillips, A. Nurduzzo, and N. E. Hussey 
for valuable discussions.

\newpage

\newpage

\newpage

\end{document}